\documentclass[prd,superscriptaddress,unsortedaddress,twocolumn,showpacs,preprintnumbers,amsmath,amssymb]{revtex4}
\usepackage[dvipdfmx]{graphicx}
\usepackage{amsmath,amssymb,times}
\usepackage{ulem}

\def\fun#1#2{\lower3.6pt\vbox{\baselineskip0pt\lineskip.9pt
\ialign{$\mathsurround=0pt#1\hfil##\hfil$\crcr#2\crcr\sim\crcr}}}

\newcommand{\beq}{\begin{equation}}
\newcommand{\eeq}{\end{equation}}
\newcommand{\bea}{\begin{eqnarray}}
\newcommand{\eea}{\end{eqnarray}}

\renewcommand{\a}{\alpha}

\newcommand{\la}{\left\langle}

\DeclareSymbolFont{boldletters}{OML}{cmm} {b}{it}
\DeclareSymbolFontAlphabet{\mathbit}{boldletters}
\DeclareMathSymbol{\alpha}{\mathalpha}{letters}{"0B}
\DeclareMathSymbol{\beta}{\mathalpha}{letters}{"0C}
\DeclareMathSymbol{\gamma}{\mathalpha}{letters}{"0D}
\DeclareMathSymbol{\delta}{\mathalpha}{letters}{"0E}
\DeclareMathSymbol{\epsilon}{\mathalpha}{letters}{"0F}
\DeclareMathSymbol{\zeta}{\mathalpha}{letters}{"10}
\DeclareMathSymbol{\eta}{\mathalpha}{letters}{"11}
\DeclareMathSymbol{\theta}{\mathalpha}{letters}{"12}
\DeclareMathSymbol{\iota}{\mathalpha}{letters}{"13}
\DeclareMathSymbol{\kappa}{\mathalpha}{letters}{"14}
\DeclareMathSymbol{\lambda}{\mathalpha}{letters}{"15}
\DeclareMathSymbol{\mu}{\mathalpha}{letters}{"16}
\DeclareMathSymbol{\nu}{\mathalpha}{letters}{"17}
\DeclareMathSymbol{\xi}{\mathalpha}{letters}{"18}
\DeclareMathSymbol{\pi}{\mathalpha}{letters}{"19}
\DeclareMathSymbol{\rho}{\mathalpha}{letters}{"1A}
\DeclareMathSymbol{\sigma}{\mathalpha}{letters}{"1B}
\DeclareMathSymbol{\tau}{\mathalpha}{letters}{"1C}
\DeclareMathSymbol{\upsilon}{\mathalpha}{letters}{"1D}
\DeclareMathSymbol{\phi}{\mathalpha}{letters}{"1E}
\DeclareMathSymbol{\chi}{\mathalpha}{letters}{"1F}
\DeclareMathSymbol{\psi}{\mathalpha}{letters}{"20}
\DeclareMathSymbol{\omega}{\mathalpha}{letters}{"21}
\DeclareMathSymbol{\varepsilon}{\mathalpha}{letters}{"22}
\DeclareMathSymbol{\vartheta}{\mathalpha}{letters}{"23}
\DeclareMathSymbol{\varpi}{\mathalpha}{letters}{"24}
\DeclareMathSymbol{\varrho}{\mathalpha}{letters}{"25}
\DeclareMathSymbol{\varsigma}{\mathalpha}{letters}{"26}
\DeclareMathSymbol{\varphi}{\mathalpha}{letters}{"27}
\DeclareMathSymbol{\Gamma}{\mathalpha}{letters}{"00}
\DeclareMathSymbol{\Delta}{\mathalpha}{letters}{"01}
\DeclareMathSymbol{\Theta}{\mathalpha}{letters}{"02}
\DeclareMathSymbol{\Lambda}{\mathalpha}{letters}{"03}
\DeclareMathSymbol{\Xi}{\mathalpha}{letters}{"04}
\DeclareMathSymbol{\Pi}{\mathalpha}{letters}{"05}
\DeclareMathSymbol{\Sigma}{\mathalpha}{letters}{"06}
\DeclareMathSymbol{\Upsilon}{\mathalpha}{letters}{"07}
\DeclareMathSymbol{\Phi}{\mathalpha}{letters}{"08}
\DeclareMathSymbol{\Psi}{\mathalpha}{letters}{"09}
\DeclareMathSymbol{\Omega}{\mathalpha}{letters}{"0A}

\def\la{\mathrel{\mathpalette\fun <}}
\def\ga{\mathrel{\mathpalette\fun >}}
\def\fun#1#2{\lower3.6pt\vbox{\baselineskip0pt\lineskip.9pt
\ialign{$\mathsurround=0pt#1\hfil##\hfil$\crcr#2\crcr\sim\crcr}}}

\begin{document}
\title{Determination of $U(1)_{\rm A}$ restoration 
from pion and $a_0$-meson screening masses: \\ Toward the chiral regime}

\author{Masahiro Ishii}
\email[]{ishii@phys.kyushu-u.ac.jp}
\affiliation{Department of Physics, Graduate School of Sciences, Kyushu University,
             Fukuoka 812-8581, Japan}             
\author{Koji Yonemura}
\email[]{yonemura@phys.kyushu-u.ac.jp}
\affiliation{Department of Physics, Graduate School of Sciences, Kyushu University,
             Fukuoka 812-8581, Japan}

\author{Junichi Takahashi}
\email[]{takahashi@phys.kyushu-u.ac.jp}
\affiliation{Department of Physics, Graduate School of Sciences, Kyushu University,
             Fukuoka 812-8581, Japan}

\author{Hiroaki Kouno}
\email[]{kounoh@cc.saga-u.ac.jp}
\affiliation{Department of Physics, Saga University,
             Saga 840-8502, Japan}  

\author{Masanobu Yahiro}
\email[]{yahiro@phys.kyushu-u.ac.jp}
\affiliation{Department of Physics, Graduate School of Sciences, Kyushu University,
             Fukuoka 812-8581, Japan}

\date{\today}

\begin{abstract} 
We incorporate the effective restoration of $U(1)_{\rm A}$ symmetry 
in the 2+1 flavor entanglement Polyakov-loop extended Nambu--Jona-Lasinio 
(EPNJL) model by introducing a temperature-dependent 
strength $K(T)$ to 
the Kobayashi-Maskawa-'t Hooft (KMT) determinant interaction. 
$T$ dependence of $K(T)$ is well determined 
from pion and $a_0$-meson screening masses obtained by 
 lattice QCD (LQCD) simulations with improved p4 staggered fermions. The strength is strongly suppressed in the vicinity of 
the pseudocritical temperature of chiral transition. The EPNJL model with the $K(T)$ well reproduces meson susceptibilities calculated by LQCD with domain-wall fermions. 
The model shows that the chiral transition is second order 
at the ``light-quark chiral-limit'' point where the light quark mass  
is zero and the strange quark mass is fixed at the physical value. 
This indicates that there exists a tricritical point. 
Hence the location is estimated.  
\end{abstract}

\maketitle

\section{Introduction}
\label{Introduction}
Meson masses are important quantities to understand the properties 
of Quantum Chromodynamics (QCD) vacuum. For example, 
the difference between pion and sigma-meson masses is mainly originated in 
the spontaneous breaking of chiral symmetry, so that the restoration 
can be determined from temperature ($T$) dependence 
of the mass difference. 
Similar analysis is possible for the effective restoration of $U(1)_{\rm A}$ 
symmetry through the difference between pion and $a_0$-meson 
(Lorentz scalar and isovector meson) masses.

$U(1)_{\rm A}$ symmetry is explicitly broken by the axial anomaly and 
the current quark mass. In the effective model,  the $U(1)_{\rm A}$ 
anomaly is simulated by the Kobayashi-Maskawa-'t Hooft (KMT) determinant 
interaction~\cite{KMK,tHooft}. The coupling constant $K$ 
of the KMT interaction is proportional 
to the instanton density screened by the medium with finite 
$T$~\cite{Pisarski-Yaffe-1980}. Hence 
$K$ becomes small as $T$ increases: $K=K(T)$. 
Pisarski and Yaffe discussed the suppression $S(T) \equiv K(T)/K(0)$ 
for high $T$, say $T \ga 2T_c$ for the pseudocritical temperature $T_c$ 
of chiral transition, 
by calculating the Debye-type screening~\cite{Pisarski-Yaffe-1980}: 
\bea
S(T)&=&
\exp \Big[-\pi^2\rho^2 T^2 \Big( \frac{2}{3}N_c+\frac{1}{3}N_f \Big) \Big]
\nonumber \\ 
&=&\exp[-T^2/b^2] ,
\label{instanton-suppression-eq}
\eea
where $N_c$ ($N_f$) is the number of colors (flavors) and 
the typical instanton radius $\rho$ is about 1/3~fm, and hence 
the suppression parameter $b$ is about $0.70 T_c$ 
for $N_c=N_f=3$ of our interest~\cite{Shuryak:1993ee}; note that 
2+1 flavor LQCD simulations show 
$T_{\rm c}= 154\pm 9$~MeV~\cite{Borsanyi:2010bp,Bazavov:2011nk,Kanaya}.
This phenomenon is called ``effective restoration of $U(1)_{\rm A}$ symmetry'', since $U(1)_{\rm A}$ symmetry is always broken in the current-operator level
but effectively restored at higher $T$ in the vacuum expectation value. 

\begin{figure}[t]
\begin{center}
   \includegraphics[width=0.5\textwidth]{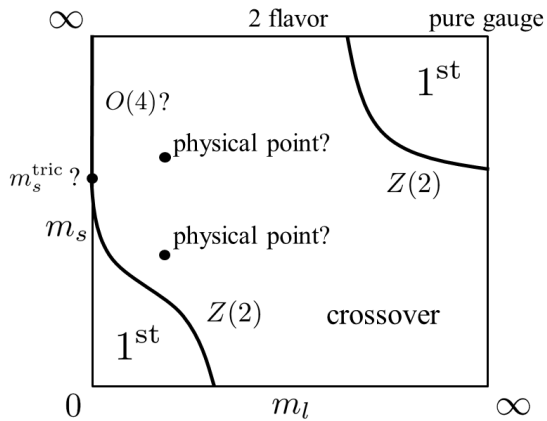}
\end{center}
\caption{A schematic phase diagram of 2+1 flavor QCD as a function of 
light-quark mass $m_l$ and strange-quark mass ${m_s}$. 
A tricritical point is likely to appear on the $ m_s$ axis; the location
 is shown by $(m_l, m_s)=(0,m_{s}^{\rm tric})$. 
The solid lines stand for second-order transitions belonging 
to the universality class labeled, 
where the labels $Z(2)$ and $O(4)$ mean the 3d Ising and 
the 3d $O(4)$ class, respectively. 
}
\label{Fig-schematic-phase-diagram}
\end{figure}

Figure \ref{Fig-schematic-phase-diagram} shows 
the current status of knowledge on 2+1 flavor phase diagram for various 
values of light-quark mass $m_l$ and strange-quark mass ${m_s}$. 
QCD shows a first-order phase transition associated with the breaking of 
chiral ($Z_3$) symmetry at the lower left (upper right) corner 
\cite{Pisarski-Wilczek-1984,Vicari:2007ma}. When $m_l$ and ${m_s}$ 
are finite, these first-order transitions become second order 
of 3d Ising  ($Z(2)$) universality class, 
as shown by the solid lines \cite{Pisarski-Wilczek-1984,Vicari:2007ma}.  

However, the order of chiral transition and its
universality class are unknown on the vertical line of $m_l=0$ and $m_s > 0$, 
and it is considered to be related 
to the effective $U(1)_{\rm A}$ restoration. 
In the two-flavor chiral limit of $(m_l,m_s)=(0,\infty)$ at 
the upper left corner, for example, the order may be second order 
of $O(4)$ class if 
the effective restoration is not completed at $T=T_c$, 
because the chiral symmetry becomes 
$SU_{\rm L}(2) \times SU_{\rm R}(2)$ isomorphic to $O(4)$ in the situation  
and the transition is 
then expected to be in the 3d $O(4)$ universality class 
\cite{Pisarski-Wilczek-1984,Vicari:2007ma}. 
When $U(1)_{\rm A}$ symmetry is restored completely at $T=T_c$, 
it was suggested in Ref. \cite{Pisarski-Wilczek-1984} that 
the chiral transition becomes the first order. 
Recently, however, it was pointed out in Ref. \cite{Vicari:2013} 
that the second order is still possible. In this case, 
the universality class is not $O(4)$ but $U_{\rm L}(2)\times U_{\rm R}(2)$.
There are many lattice QCD (LQCD) simulations made so far
to clarify the order and its universality class in 
the two-flavor chiral limit of $(m_l,m_s)=(0,\infty)$ 
and the light-quark chiral limit 
where $m_l$ vanishes with $m_s$ fixed at the physical value, 
but these are still controversial; see Refs. 
\cite{Fukugita-PRL1990,Fukugita-PRD1990,Brown-1990,Karsch-1994-1,Karsch-1994-2,Iwasaki-1997,Aoki-1998,Khan-2000,Bernard-2000,D'Elia-2005,Ejiri-2009,Bonati:2014kpa} 
and therein.

Very recently, the effective restoration of $U(1)_{\rm A}$ symmetry was 
investigated by pion and $a_0$-meson screening masses calculated with 
LQCD with improved p4 staggered 
fermions~\cite{Cheng:2010fe} and also by 
meson susceptibilities calculated with LQCD with 
domain-wall fermions~\cite{Bhattacharya:2014ara,Buchoff:2013nra}. 
The effective restoration of $U(1)_{\rm A}$ symmetry thus becomes 
an important current issue.

In LQCD, pole and screening masses are evaluated from 
the exponential decay of mesonic correlation functions in the temporal and 
spatial directions, respectively, but for finite $T$ the lattice size is 
smaller in the temporal direction than in the spatial direction. 
This makes LQCD simulations less feasible for pole masses 
than for screening masses. The problem is getting serious as $T$ increases. 
This is the reason why meson screening masses are calculated 
in most of LQCD simulations. 
In fact, as mentioned above, state-of-the-art LQCD calculations were done 
for meson screening masses with large volumes 
($16^3 \times 4$, $24^3 \times 6$, $32^3 \times 8$) in a wide range of 
$T$~\cite{Cheng:2010fe}.

The effective model is suitable for qualitative understanding of QCD. 
In fact, the QCD phase diagram, the properties of light-meson 
pole masses and the $T$ dependence of $U(1)_{\rm A}$ anomaly
have been studied extensively with 
the Polyakov-loop extended Nambu--Jona-Lasinio (PNJL) model
~\cite{Meisinger,Dumitru,Costa:2005,Fukushima1,Ghos,Megias,Ratti1,Ciminale,Ratti2,Rossner,Hansen,Sasaki-C,Schaefer,Kashiwa1,Costa:2009,Ruivo:2012a}. 
The PNJL model can treat the deconfinement and chiral transitions 
simultaneously, but cannot reproduce their coincidence seen in 
2-flavor LQCD data, when the model parameters are properly set 
to reproduce $T_c$ calculated with LQCD~\cite{Ratti1}. 
This problem is solved by introducing the four-quark vertex depending 
on the Polyakov-loop~\cite{Sakai_EPNJL,Sasaki_EPNJL}. 
The model with the entanglement vertex is refer to as 
the entanglement-PNJL (EPNJL) model. 
The EPNJL model is successful in reproducing the phase diagram 
in 2-flavor QCD at imaginary chemical potential~\cite{D'Elia3,FP2010} and 
real isospin chemical potential~\cite{Kogut2}, and well accounts for 
the phase diagram in the $m_l$--$m_s$ plane, i.e., the so-called Columbia plot 
in 2+1 flavor QCD~\cite{FP2007}. 
So far, $T$ dependence of $U(1)_{\rm A}$ anomaly and low-lying meson pole masses was 
studied extensively by both the PNJL and EPNJL 
models~\cite{Costa:2005,Costa:2009,Ruivo:2012a,Ruivo:2012b}, but 
$T$ dependence of meson screening masses was investigated 
by the EPNJL model only in our previous work~\cite{Ishii:2013kaa}.

In NJL-type models, it is well known that the calculation of 
meson screening mass $M_{\xi,{\rm scr}}$ is difficult, 
where $\xi$ denotes the type of meson. 
In fact, only a few trials were made so far~\cite{Kunihiro,Florkowski}. 
The first problem is the regularization. The regularization widely 
used is the three-dimensional momentum cutoff, but it breaks 
Lorentz invariance at $T=0$ and spatial-translation invariance at any $T$. 
This generates 
unphysical oscillations in the spatial correlation function 
$\eta_{\xi\xi}(r)$ ~\cite{Florkowski}. 
This refuses us to determine $M_{\xi,{\rm scr}}$ from 
the asymptotic form of $\eta_{\xi\xi}(r)$ as 
\begin{equation}
M_{\xi,{\rm scr}}
=-\lim_{r\rightarrow \infty}\frac{d \ln{\eta_{\xi\xi}(r)}}{dr}. 
\label{scr-mass}
\end{equation}
This problem can be solved~\cite{Florkowski} by introducing 
the Lorentz-invariant Pauli-Villars (PV) regularization~\cite{PV}.

Even after the unphysical oscillations are removed, 
heavy numerical calculations 
are still required to obtain $\eta_{\xi\xi}(r)$ 
at large $r$~\cite{Florkowski}. This is the second problem. 
In the model calculation, the spatial correlation function is obtained 
first in the momentum representation $(\tilde{q}=\pm |{\bf q}|)$ 
as $\chi_{\xi\xi}(0, \tilde{q}^{2})$. Hence we have to make 
the Fourier transform 
from $\chi_{\xi\xi}(0, \tilde{q}^{2})$ to $\eta_{\xi\xi}(r)$: 
\begin{equation}
\eta_{\xi\xi}(r)=\frac{1}{4\pi^{2}ir}\int^{\infty}_{-\infty}d\tilde{q}\hspace{1ex}\tilde{q}\chi_{\xi\xi}(0,\tilde{q}^2)e^{i\tilde{q}r} .
\label{chi_r}
\end{equation}
The $\tilde{q}$ integration is quite hard particularly at large $r$, 
since the integrand consists of a slowly damping function 
$\tilde{q}\chi_{\xi\xi}(0,\tilde{q}^2)$ and a highly oscillating function 
$e^{i\tilde{q}r}$. 
If $\chi_{\xi\xi}(0, \tilde{q}^{2})$ 
has a pole below the cut in the complex $\tilde{q}$ plane, 
one can easily determine $M_{\xi,{\rm scr}}$ from the pole location. 
In the old formulation of Ref.~\cite{Florkowski}, 
the condition was not satisfied, since logarithmic cuts appear 
in the vicinity of the real $\tilde{q}$ axis in addition to physical cuts. 
Very recently we solved the problem 
in our previous paper~\cite{Ishii:2013kaa}, showing that 
the logarithmic cuts near the real $\tilde{q}$ axis are 
unphysical and removable. 
In the new formulation based on the PV regularization, 
there is no logarithmic cut and 
a pole appears below physical cuts, as shown later.

In this paper, 
we incorporate the effective restoration of $U(1)_{\rm A}$ symmetry 
in the 2+1 flavor EPNJL model 
by introducing a $T$-dependent coupling strength $K(T)$ 
to the KMT interaction. 
$T$ dependence of $K(T)$ is well determined 
from state-of-the-art 2+1 flavor LQCD results~\cite{Cheng:2010fe} 
on pion and $a_0$-meson screening masses. 
For the derivation of meson screening mass, we extend the previous 
prescription of Ref.~\cite{Ishii:2013kaa} for 2 flavors to 2+1 flavors. 
The $K(T)$ determined from the LQCD data 
is strongly suppressed near $T_c$. Using the parameter set, we show that 
the chiral transition is second order in  
the light-quark chiral limit.  
This result indicates that there exists a tricritical point 
near the ``light-quark chiral-limit'' point in the $m_l$--$m_s$ plane. 
We then estimate the location.

We recapitulate the EPNJL model and the method of 
calculating meson screening masses in Sec. \ref{Model setting} and 
show the results of numerical calculations in Sec. \ref{Numerical Results}.  
Section \ref{Summary} is devoted to a summary.

\section{Model setting}
\label{Model setting}

\subsection{EPNJL model}
\label{EPNJL model}

We start with the 2+1 flavor EPNJL model~\cite{Sakai_EPNJL,Sasaki_EPNJL}. 
The Lagrangian density is  
\begin{align}
 {\cal L}  
=& {\bar \psi}(i \gamma_\nu D^\nu - {\hat m_0} )\psi  
  + G_{\rm s}(\Phi) \sum_{a=0}^{8} 
    [({\bar \psi} \lambda_a \psi )^2 +({\bar \psi }i\gamma_5 \lambda_a \psi )^2] 
\nonumber\\
 &- K(T) \Bigl[\det_{f,f'} {\bar \psi}_f (1+\gamma_5) \psi_{f'} 
           +\det_{f,f'} {\bar \psi}_f (1-\gamma_5) \psi_{f'} \Bigr]
\nonumber\\
&-{\cal U}(\Phi [A],{\bar \Phi} [A],T) 
\label{L}
\end{align} 
with quark fields $\psi=(\psi_u,\psi_d,\psi_s)^T$ and 
$D^\nu=\partial^\nu + iA^\nu$ with $A^\nu=\delta^{\nu}_{0}g(A^0)_a{t_a / 2}=-\delta^{\nu}_{0}ig(A_{4})_a t_a/2$ for the gauge coupling
$g$, where the $\lambda_a$ ($t_a$) are the Gell-Mann matrices 
in flavor (color) space and $\lambda _0 = \sqrt{2/3}~{\bf I}$ 
for the unit matrix ${\bf I}$ in flavor space. The determinant in
\eqref{L} is taken in flavor space.
For the 2+1 flavor system, the current quark masses 
${\hat m_0}={\rm diag}(m_u,m_d,m_s)$ satisfy a relation 
$m_s > m_l \equiv m_u=m_d$. 
In the EPNJL model, the coupling strength $G_{\rm s}(\Phi)$ 
of the scalar-type four-quark interaction depends 
on the Polyakov loop $\Phi$ and its Hermitian conjugate 
${\bar \Phi}$ as 
\begin{equation}
G_{\rm s}(\Phi)=G_{\rm s}(0)
\left[
1-\alpha_1\Phi{\bar \Phi}-\alpha_2 (\Phi^3+{\bar \Phi}^3) 
\right]. 
\label{EPNJL}
\end{equation}
This entanglement coupling is charge-conjugation and $Z_{3}$ symmetric. 
When $\alpha_1=\alpha_2=0$, the EPNJL model is 
reduced to the PNJL model. 
We set $\alpha_2=0$ for simplicity, since 
the $\alpha_2$ term yields the same effect as the $\alpha_1$ term 
in the present analysis. 
As shown later in Sec. \ref{Numerical Results}, 
the value of $\a_1$ is determined 
from LQCD data on pion and $a_0$-meson screening masses; 
the resulting value is $\a_1 = 1.0$.

For $T$ dependence of $K(T)$, we assume the following form phenomenologically: 
 \begin{eqnarray}
 K(T)=\left\{ \begin{array}{ll}
 K(0) & (T < T_1) \\
 K(0) e^{-(T-T_1)^2/b^2}& (T \ge T_1) \\
 \end{array} \right. . 
 \label{T-dependent-K}
 \end{eqnarray}
For high $T$ satisfying $T \gg T_1$, the form  \eqref{T-dependent-K} 
is reduced to \eqref{instanton-suppression-eq}. 
As shown later in Sec. \ref{Numerical Results}, the values of 
$T_1$ and $b$ are well determined 
from LQCD data on pion and $a_0$-meson screening masses; 
the resulting values are 
$T_1=0.79T_{c}=121$~MeV and $b = 0.23T_c = 36$~MeV. 

After the Pisarski-Yaffe discussion on $S(T)$, 
$T$ dependence of the instanton density was estimated theoretically by the 
the instanton-liquid model \cite{Shuryak:1993ee}, 
but the estimation is applicable only for $T \ga 2T_c$. For this reason, 
in Ref.~\cite{Ruivo:2012b}, a Woods-Saxon form $(1+e^{(T-T_1')/b'})^{-1}$ 
with two parameters $T_1'$ and $b'$ was used phenomenologically 
for $K(T)/K(0)$. The present form \eqref{T-dependent-K} has $T$ dependence 
similar to the Woods-Saxon form.

In the EPNJL model, the time component of $A_\mu$ is treated as a homogeneous 
and static background field, which is governed by the Polyakov-loop potential 
$\mathcal{U}$. In the Polyakov gauge, $\Phi$ and ${\bar \Phi}$ 
are obtained  by  
\begin{align}
\Phi &= {1\over{3}}{\rm tr}_{\rm c}(L),
~~~~~{\bar \Phi} ={1\over{3}}{\rm tr}_{\rm c}({L^*})
\label{Polyakov}
\end{align}
with $L= \exp[i A_4/T]=\exp[i {\rm diag}(A_4^{11},A_4^{22},A_4^{33})/T]$ 
for real variables $A_4^{jj}$ 
satisfying $A_4^{11}+A_4^{22}+A_4^{33}=0$. 
For zero quark chemical potential where $\Phi={\bar \Phi}$, 
one can set $A^{33}_4=0$ and determine the others as 
$A^{22}_4=-A^{11}_4={\rm cos}^{-1}[(3\Phi -1)/2]$.

We use the logarithm-type Polyakov-loop potential of Ref.~\cite{Rossner} 
as $\mathcal{U}$. The parameter set in $\mathcal{U}$ has already been 
determined from  LQCD data at finite $T$ in the pure gauge limit. 
The potential has a parameter $T_0$ and yields a first-order deconfinement 
phase transition at $T=T_0$. The parameter used to be set to $T_0=270$~MeV, 
since LQCD data show the phase transition at $T=270$~MeV in 
the pure gauge limit. In full QCD with dynamical quarks, however, 
the PNJL model with this value of $T_0$ is found not to explain LQCD 
results. Nowadays, $T_0$ is then rescaled to reproduce the LQCD results. 
In the present case, we take $T_0=180$~MeV so that the EPNJL model 
can reproduce LQCD results for the pseudocritical temperature 
$T_c^{\rm deconf}$ of deconfinement transition; 
actually, $T_c^{\rm deconf}=165$ MeV in the EPNJL model and 
$170 \pm 7$ MeV in LQCD \cite{Aoki:2009sc}.

Making the mean field approximation (MFA) to \eqref{L} leads to 
the linearized Lagrangian density 
\begin{eqnarray}
  {\cal L}^{\rm MFA}  
= {\bar \psi}S^{-1}\psi  - U_{\rm M} - {\cal U}(\Phi [A],{\bar \Phi} [A],T)
\label{linear-L}
\end{eqnarray}
with the quark propagator 
\bea
S=(i \gamma_\nu \partial^\nu - \gamma_0A^0 - \hat{M})^{-1},
\eea
where $\hat{M}={\rm diag}(M_u,M_d,M_s)$ with 
\begin{eqnarray*}
M_u &=& m_u -4G_{\rm s}(\Phi)\sigma_u +2K(T)\sigma_d \sigma_s ,\\
M_d &=& m_d -4G_{\rm s}(\Phi)\sigma_d +2K(T)\sigma_s \sigma_u ,\\
M_s &=& m_s -4G_{\rm s}(\Phi)\sigma_s +2K(T)\sigma_u \sigma_d ,
\end{eqnarray*}
and $\sigma_f $ means the chiral condensate $\langle \bar
\psi_f \psi_f\rangle$ for flavor $f$. The mesonic potential $U_{\rm M}$ is 
\begin{eqnarray*}
U_{\rm M}
= 2G_{\rm s}(\Phi)(\sigma^2_u+\sigma^2_d+\sigma^2_s)  
-4K(T) \sigma_u \sigma_d \sigma_s.
\end{eqnarray*}
Making the path integral over quark fields, one can get 
the thermodynamic potential (per unit volume) as
\begin{align}
&\Omega= U_{\rm M}+{\cal U}-2 \sum_{f=u,d,s} \int \frac{d^3 {\bf p}}{(2\pi)^3}
   \Bigl[ 3 E_{p,f} \notag \\
&+ \frac{1}{\beta}
           \ln~ [1 + 3(\Phi+{\bar \Phi} e^{-\beta E_{p,f})} 
           e^{-\beta E_{p,f}}+ e^{-3\beta E_{p,f}}] \notag\\
&+ \frac{1}{\beta} 
           \ln~ [1 + 3({\bar \Phi}+{\Phi e^{-\beta E_{p,f}}}) 
              e^{-\beta E_{p,f}}+ e^{-3\beta E_{p,f}}]
	      \Bigl]
\label{PNJL-Omega}
\end{align}
with 
$E_{p,f}=\sqrt{{\bf p}^2+M^2_f}$ and $\beta = 1/T$. We determine the mean-field variables 
($X=\sigma_l,\sigma_s,\Phi,\bar{\Phi}$) 
from the stationary conditions: 
\begin{equation}
\frac{\partial \Omega}{\partial X}=0,
\end{equation}
where isospin symmetry is assumed for the light-quark sector, 
i.e., $\sigma_l \equiv \sigma_u=\sigma_d$.

On the right-hand side of (\ref{PNJL-Omega}), the first term (vacuum term) 
in the momentum integral diverges. We then use 
the PV regularization~\cite{Florkowski,PV}. 
In the scheme, the integral $I(M_f,q)$ is regularized as 
\begin{eqnarray}
I^{\rm reg}(M_f,q)=\sum_{\alpha=0}^2 C_\alpha I(M_{f;\alpha},q) ,
\label{PV}
\end{eqnarray}
where $M_{f;0}=M_f$ and the $M_{f;\alpha}~(\alpha\ge 1)$ mean masses 
of auxiliary particles. The parameters $M_{f;\alpha}$ and $C_\alpha$ 
should satisfy the condition  
$\sum_{\alpha=0}^2C_\alpha=\sum_{\alpha=0}^2 C_\alpha M_{f;\alpha}^2=0$.
We then assume $(C_0,C_1,C_2)=(1,1,-2)$ and 
$(M_{f;1}^2,M_{f;2}^2)=(M^2_{f}+2\Lambda^2,M^2_{f}+\Lambda^2)$. 
We keep the parameter $\Lambda$ finite 
even after the subtraction \eqref{PV}, since the present model is 
non-renormalizable. 
The parameters are taken from Ref. \cite{Bernard} and they are
$m_l=6.2$ MeV, $m_s=175.0$ MeV, $G_{\rm s}(0)\Lambda^2=2.35$ and 
$K(0)\Lambda^5=27.8$ for $\Lambda =795$ MeV. 
This parameter set reproduces mesonic observables at
vacuum, i.e., the pion and kaon decay constants 
($f_{\pi}=92$ MeV and $f_K=105$ MeV) and their masses 
($M_{\pi}=141$ MeV and $M_K=512$ MeV) and the $\eta'$-meson mass 
($M_{\eta'}=920$ MeV).
In the present work, we analyze LQCD results
of Ref. \cite{Cheng:2010fe} for pion and $a_0$-meson screening masses. 
In the LQCD simulation, the pion mass $M_{\pi}(0)$ at vacuum ($T=0$) 
is 175~MeV and a bit heavier than the experimental value 138~MeV. 
We then change $m_l$ to $9.9$~MeV in the EPNJL model in order to reproduce 
$M_{\pi}(0)=175~{\rm MeV}$. This parameter set yields
$M_{a_0}(0)=711$ MeV, $M_{\eta}(0)=481$ MeV and $M_{\sigma}(0)=537$ MeV
as $a_0,\eta$ and $\sigma$ meson pole masses at vacuum.

\subsection{Meson pole mass}
We derive the equations for pion and $a_0$-meson pole masses, 
following Ref~\cite{Hansen,KHKB}. 
The current corresponding to a meson of type $\xi$ is 
\beq
  J_{\xi}(x) = \bar \psi(x)\Gamma_\xi \psi(x)
             - \langle\bar \psi(x) \Gamma_\xi \psi(x)\rangle , 
\label{source}
\eeq
where 
$\Gamma_\pi=i\gamma_5\lambda_3$ for $\pi$ meson and 
$\Gamma_{a_0}=\lambda_3$ for $a_0$-meson. 
We denote the Fourier transform of the mesonic correlation function 
$\eta_{\xi\xi} (x) \equiv \langle 0 | {\rm T} \left( J_\xi(x) J^{\dagger}_{\xi}(0) \right) | 0 \rangle$ by $\chi_{\xi\xi} (q^2)$ as 
\begin{align}
\chi_{\xi\xi} (q^2) = \chi_{\xi\xi} (q^2_0, {\tilde q}^2) 
=  i \int d^4x e^{i q\cdot x} \eta_{\xi\xi} (x) ,
\end{align}
where $\tilde{q}=\pm |{\bf q}|$ for $q=(q_0,{\bf q})$ and 
${\rm T}$ stands for the time-ordered product. 
Using the random-phase (ring) approximation,
one can obtain the Schwinger-Dyson equation
\bea
  \chi_{\xi\xi}
= \Pi_{\xi\xi} +2 \sum_{\xi'\xi''} \Pi_{\xi\xi'}G_{\xi'\xi''}\chi_{\xi''\xi} 
\label{SD}
\eea
for $\chi_{\xi\xi}$, where 
$G_{\xi'\xi''}$ is an effective four-quark interaction and 
$\Pi_{\xi\xi'}$ is the one-loop polarization function defined by 
\bea
  \Pi_{\xi\xi'}(q^2)  \equiv  (-i) \int \frac{d^4 p}{(2\pi)^4} 
  {\rm Tr} \left( \Gamma_\xi iS(p'+q) \Gamma_{\xi'} iS(p') \right) 
\eea
with $p'=(p_{0}+iA_4,{\bf p})$, where the trace ${\rm Tr}$ is taken 
in flavor, Dirac and 
color spaces. Here the quark propagator $S(p)$ in momentum space 
is diagonal in flavor space: $S(p)={\rm diag}(S_u,S_d,S_s)$. 
For $\xi=\pi$ and $a_0$, furthermore, $G_{\xi\xi'}$
and $\Pi_{\xi\xi'}$ are diagonal 
($G_{\xi\xi'}=G_{\xi}\delta_{\xi\xi'}$, $\Pi_{\xi\xi'}=\Pi_{\xi}\delta_{\xi\xi'}$), because we impose isospin symmetry for the light-quark sector and 
employ the random-phase approximation. One can then easily 
solve the Schwinger-Dyson equation for $\xi=\pi$ and $a_0$: 
\bea
\chi_{\xi\xi} &=& 
\frac{\Pi_{\xi}}{1 - 2G_{\xi}\Pi_{\xi}}
\label{chi_pia0}  
\eea
with the effective couplings $G_{\pi}$ and $G_{a_0}$ defined by 
\begin{eqnarray}
G_{a_0} &=& G_{\rm s}(\Phi) + \frac{1}{2}K(T) \sigma_s, 
\label{g_a0}
\\
G_{\pi} &=& G_{\rm s}(\Phi) - \frac{1}{2}K(T) \sigma_s.
\label{g_pi}
\end{eqnarray}

As for $T=0$,  $\Pi_{\pi}$ and $\Pi_{a_0}$ have the following explicit forms: 
\begin{eqnarray}
\Pi_{a_0}&=&i\sum_{f,f'}(\lambda_3)_{f'f}(\lambda_3)_{ff'}\nonumber\\
&\times&
\int {d^4p\over{(2\pi )^4}}{\rm tr}_{\rm c,d}\Bigl[{\{\gamma_\mu (p'+q)^\mu +M_f\}(\gamma_\nu p'^\nu +M_{f'})\over{\{(p'+q)^2-M_f^2\}(p'^2-M_{f'}^2)}}\Bigr]
\nonumber\\
&=&4i[I_{1}+I_{2}-(q^2-4M^2)I_{3}], 
\label{Pi_SS}
\\
\Pi_{\pi}&=&i\sum_{f,f'}(\lambda_3)_{f'f}(\lambda_3)_{ff'}\nonumber\\
&\times&\int {d^4p\over{(2\pi )^4}}{\rm tr}_{\rm c,d}\Bigl[(i\gamma_5){\{\gamma_\mu (p'+q)^\mu +M_f\}\over{\{(p'+q)^2-M_f^2\}}}
\nonumber\\
&&\times (i\gamma_5)
{(\gamma_\nu p'^\nu +M_{f'})\over{(p'^2-M_{f'})^2}}\Bigr]
\nonumber\\
&=&4i[I_{1}+I_{2}-q^2I_{3}], 
\label{Pi_PP}
\end{eqnarray}
and 
\begin{eqnarray}
I_{1}&=&\int {d^4p\over{(2\pi )^4}}{\rm tr_c}\Bigl[{1\over{p'^2-M}}\Bigr],
\label{I1}
\\
I_{2}&=&\int {d^4p\over{(2\pi )^4}}{\rm tr_c}\Bigl[{1\over{(p'+q)^2-M^2}}\Bigr],
\label{I2}
\\
I_{3}&=&\int {d^4p\over{(2\pi )^4}}{\rm tr_c}\Bigl[{1\over{\{(p'+q)^2-M^2\}(p'^2-M^2)}}\Bigr] ,
\nonumber\\
\label{I3}
\end{eqnarray}
where ${\rm tr}_{\rm c,d}$ (${\rm tr}_{\rm c}$) means 
the trace in color and Dirac spaces (color space) and $M=M_u=M_d$. 
For finite $T$, the corresponding equations are 
obtained by the replacement
\begin{align}
&p_0 \to i \omega_n = i(2n+1) \pi T, 
\nonumber\\
&\int \frac{d^4p}{(2 \pi)^4} 
\to iT\sum_{n=-\infty}^{\infty} \int \frac{d^3{\bf p}}{(2 \pi)^3}. 
\label{finte_T_mu}
\end{align}

The meson pole mass $M_{\xi,{\rm pole}}$ is a pole of 
$\chi_{\xi\xi}(q_0^2,{\tilde q}^2)$ in the complex $q_0$ plane. 
Taking the rest frame $q=(q_0,{\bf 0})$ for convenience, 
one can get the equation for $M_{\xi,{\rm pole}}$ as 
\begin{align}
\big[1 - 2G_{\xi}\Pi_{\xi}(q_0^2,0)\big]\big|_{q_0=M_{\xi,{\rm pole}}-i\frac{\Gamma}{2}}=0 ,   
\label{mmf}
\end{align}
where $\Gamma$ is the decay width to $q\bar{q}$ continuum. The method of
calculating meson pole masses is well established in the PNJL
model~\cite{Hansen, KHKB}. 

\subsection{Meson screening mass}
We derive the equations for pion and $a_0$-meson screening masses, following 
Ref.~\cite{Ishii:2013kaa}. 
This is an extension of the method of Ref.~\cite{Ishii:2013kaa} 
for 2 flavors to 2+1 flavors.

As mentioned in Sec. \ref{Introduction}, it is not easy to 
make the Fourier transform from $\chi_{\xi\xi} (0,\tilde{q}^{2})$ 
to $\eta_{\xi\xi}(r)$ particularly at large $r$. 
When the direct integration on the real $\tilde{q}$ axis is difficult, 
one can consider a contour integral in the complex $\tilde{q}$ plane 
by using the Cauchy's integral theorem. 
However, $\chi_{\xi\xi} (0,\tilde{q}^{2})$ has logarithmic cuts 
in the vicinity of the real $\tilde{q}$ axis \cite{Florkowski}, 
and it is reported in Ref. \cite{Florkowski} that 
heavy numerical calculations are necessary for evaluating the cut effects. 
In our previous work~\cite{Ishii:2013kaa}, we showed 
that these logarithmic cuts are unphysical and removable. Actually, we have no logarithmic cut, when analytic continuation is made 
for the $I_3({\bf q})$ after ${\bf p}$ integration. 
Namely, the Matsubara summation over $n$ should be taken 
after the ${\bf p}$ integration in \eqref{finte_T_mu}. 
We then express 
$I_3^{\rm reg}$ as an infinite series of analytic functions: 
\begin{eqnarray}
&&I_{3}^{\rm reg}(0,\tilde{q}^{2})=iT\sum_{j=1}^{N_c}\sum_{n=-\infty}^\infty\sum_{\alpha=0}^2C_{\alpha}
\nonumber\\
&\times & \int {d^3{\bf p}\over{(2\pi )^3}}
\Bigl[{1\over{{\bf p}^2+M_{j,n,\alpha}^2}}{1\over{({\bf p}+{\bf q})^2+M_{j,n,\alpha}^2}}\Bigr]\nonumber\\
&=&{iT\over{2\pi^2}}\sum_{j,n,\alpha}C_{\alpha}\int_0^1dx\int_0^\infty dk{k^2\over{[k^2+(x-x^2)\tilde{q}^2+M_{j,n,\alpha}^2]^2}}
\nonumber\\
&=&{iT\over{4\pi \tilde{q}}}\sum_{j,n,\alpha}C_{\alpha}  \sin^{-1}{\Bigl({{\tilde{q}\over{2}}\over{\sqrt{{\tilde{q}^2\over{4}}+M_{j,n,\alpha}^2}}}\Bigr)} 
\label{I_3_final}
\end{eqnarray}
with 
\begin{equation}
M_{j,n,\alpha}(T)=\sqrt{M_{\alpha}^2 + \{(2n+1)\pi T+A_{4}^{jj}\}^2 }, 
\label{KK_mode}
\end{equation}
where $M_\alpha=M_{u;\alpha}=M_{d;\alpha}$. We have numerically confirmed 
that the convergence of $n$-summation is quite fast in (\ref{I_3_final}). 
Each term of $I_3^{\rm reg}(0,\tilde{q}^2)$ 
has  two physical cuts on the imaginary axis, one is an upward vertical line 
starting from $\tilde{q} = 2iM_{j,n,\alpha}$ and the other is 
a downward vertical line from $\tilde{q} = - 2iM_{j,n,\alpha}$. 
The lowest branch point is $\tilde{q}=2iM_{j=1,n=0,\alpha=0}$. 
The value is the meson screening mass in which there is no interaction 
between a quark and an antiquark, i.e., $G_{\xi}=0$. 
Hence we may call $2M_{j=1,n=0,\alpha=0}$ ``the threshold mass''. 

We can obtain the meson screening mass $M_{\xi,{\rm scr}}$ as a pole of
$\chi_{\xi\xi}(0,\tilde{q}^{2})$,
\begin{align}
\big[1 - 2G_{\xi} \Pi_{\xi}(0,\tilde{q}^2)\big]\big|_{\tilde{q}=iM_{\xi, {\rm scr}}}=0  .  
\label{meson_screening}
\end{align}
If the pole at $\tilde{q}=iM_{\xi, {\rm scr}}$ is well isolated from the cut, 
i.e., $M_{\xi,{\rm scr}}<2M_{j=1,n=0,\alpha=0}$, one can determine 
the screening mass from the pole location 
without making the $\tilde{q}$ integral. 
In the high-$T$ limit, the condition tends to $M_{\xi,{\rm scr}}<2 \pi T$. 

\subsection{Meson susceptibility}
We consider meson susceptibilities 
$\chi_{\xi}^{\rm sus}$ for $\xi=\pi,a_0,\eta$ and $\sigma$. 
In LQCD simulations of Refs. \cite{Bhattacharya:2014ara,Buchoff:2013nra}, 
the $\chi_{\xi}^{\rm sus}$ are defined 
in Euclidean spacetime $x_{\rm E}=(\tau,{\bf x})$ as 
\begin{eqnarray}
\chi_{\xi}^{\rm sus} 
= \frac{1}{2}\int d^{4}x_{\rm E}~
\langle J_{\xi}(\tau,{\bf x})J_{\xi}^\dagger(0,{\bf 0})\rangle .
\end{eqnarray}
In the simulations, $J_\sigma$ and $J_\eta$ are assumed to have no 
s-quark component for simplicity: namely, 
$J_\sigma=\sum_{f=u,d}\bar{\psi}_f\psi_f - \langle\sum_{f=u,d}\bar{\psi}_f\psi_f\rangle$ and $J_\eta=\sum_{f=u,d}\bar{\psi}_fi\gamma_5\psi_f -\langle\sum_{f=u,d}\bar{\psi}_fi\gamma_5\psi_f\rangle$. 
For consistency, we take the same assumption also in the present 
analysis, and denote the mesons with no s-quark (ns) component by 
$\sigma_{\rm ns}$ and $\eta_{\rm ns}$. 
The factor $1/2$ is introduced to define the $\chi_{\xi}^{\rm sus}$ 
as single-flavor quantities.

The $\chi_{\xi}^{\rm sus}$ is related to the 
the Matsubara Green's function $\chi_{\xi\xi}^{\rm E}(q_4^2,{\bf q}^2)$ 
in the momentum representation as 
\begin{eqnarray}
\chi_{\xi}^{\rm sus} = \frac{1}{2}\chi_{\xi\xi}^{\rm E}(q_4^2,{\bf q}^2)
\big|_{q_4=0, {\bf q}=0} , 
\end{eqnarray}
and $\chi_{\xi\xi}^{\rm E}$ is obtainable from \eqref{chi_pia0} 
for $\pi$ and $a_0$ mesons. For $\eta_{\rm ns}$ meson, we have to consider 
a mixing between $\eta_{\rm ns}$ and 
$\eta_{\rm s}=\bar{\psi}_si\gamma_5\psi_s$. As a result, one can 
obtain $\chi_{\eta_{\rm ns}\eta_{\rm ns}}$ as \cite{KHKB} 
\begin{equation}
\chi_{\eta_{\rm ns}\eta_{\rm ns}}=\frac{(1-2G_{\eta_{\rm s}\eta_{\rm
 s}}\Pi_{\eta_{\rm s}\eta_{\rm s}})\Pi_{\eta_{\rm ns}\eta_{\rm
 ns}}}
{\det{\left[{\boldsymbol I} - 2{\boldsymbol G}{\boldsymbol
       \Pi}\right]}},
\label{chi_eta}
\end{equation} 
where ${\bf I}$ is the unit matrix and 
\begin{eqnarray}
{\boldsymbol G}
=
\left(
\begin{array}{cc}
G_{\eta_{\rm s}\eta_{\rm s}}& G_{\eta_{\rm s}\eta_{\rm ns}}\\
G_{\eta_{\rm ns}\eta_{\rm s}}& G_{\eta_{\rm ns}\eta_{\rm ns}}
\end{array}
\right),~~
{\boldsymbol \Pi}
=
\left(
\begin{array}{cc}
\Pi_{\eta_{\rm s}\eta_{\rm s}}& 0\\
0& \Pi_{\eta_{\rm ns}\eta_{\rm ns}}
\end{array}
\right)~~ 
\label{gpi_eta}
\end{eqnarray}
for the elements 
\begin{eqnarray}
G_{\eta_{\rm s}\eta_{\rm s}}   &=& G_{\rm s}(\Phi) , \\
\label{gelement_eta1}
G_{\eta_{\rm ns}\eta_{\rm ns}} &=& G_{\rm s}(\Phi) + \frac{1}{2}K(T)\sigma_{s}, \\
\label{gelement_eta2}
G_{\eta_{\rm s}\eta_{\rm ns}}  &=& G_{\eta_{\rm ns}\eta_{\rm s}} = \frac{\sqrt{2}}{2}K(T)\sigma_l.
\label{gelement_eta3}
\end{eqnarray}
In the isospin symmetric case we consider, the polarization functions 
$\Pi_{\eta_{\rm s}\eta_{\rm s}}$ and $\Pi_{\eta_{\rm ns}\eta_{\rm ns}}$ 
have the same function form as $\Pi_{\pi}$: 
\begin{eqnarray}
\Pi_{\eta_{\rm s}\eta_{\rm s}}  &=& \Pi_{\pi}(M_{s}), \\
\label{pielement_eta1}
\Pi_{\eta_{\rm ns}\eta_{\rm ns}}&=& \Pi_{\pi}(M),  
\label{pielement_eta2}
\end{eqnarray}
where note that $\Pi_{\pi}(M_{s})$ is a function of not $M$ but $M_s$. 
Similarly, $\chi_{\sigma_{\rm ns}\sigma_{\rm ns}}$ is obtainable from 
(\ref{chi_eta}) with $K(T)$ replaced by $-K(T)$ and $\Pi_{\pi}$ by 
$\Pi_{a_0}$.  

\section{Numerical Results}
\label{Numerical Results}

\subsection{Meson screening masses}
\label{Meson screening masses}

The EPNJL model has three adjustable parameters, 
$\a_1$ in the entanglement coupling $G_{\rm s}(\Phi)$ and 
$b$ and $T_1$ in the KMT interaction $K(T)$.  
These parameters can be clearly determined from 
LQCD data~\cite{Cheng:2010fe} for 
pion and $a_0$-meson screening masses, 
$M_{\pi,{\rm scr}}$ and $M_{a_0,{\rm scr}}$, as shown below. 

Figure~\ref{Fig-screening} shows $T$ dependence of 
$M_{\pi,{\rm scr}}$ and $M_{a_0,{\rm scr}}$. 
Best fitting is obtained, when 
$\a_1=1.0$, $T_1=0.79T_{c}=121$~MeV and $b = 0.23T_c = 36$~MeV. 
Actually, the EPNJL results (solid and dot-dash lines) 
with this parameter set well account for LQCD data~\cite{Cheng:2010fe} 
for both $M_{\pi,{\rm scr}}$ and $M_{a_0,{\rm scr}}$. 
The parameters thus obtained indicate the strong suppression of $K(T)$ in the vicinity of $T_c$.  
The mass difference 
$\Delta M_{\rm scr}(T)=M_{a_0,{\rm scr}}(T)-M_{\pi,{\rm scr}}(T)$ 
is sensitive to $K(T)$ because of \eqref{g_a0} and 
\eqref{g_pi}, and hence 
the values of $b$ and $T_1$ are well determined from $\Delta M_{\rm scr}(T)$.

When $\a_1=0$, the EPNJL model is reduced to the PNJL model. 
The results of the PNJL model are 
shown in Fig.~\ref{Fig-screening-K(T)-PNJL} for comparison. 
The PNJL results cannot reproduce LQCD data particularly in the region 
$T \ga 180$~MeV. The slope of the solid and dot-dash lines in the region 
is thus sensitive to the value of $\a_1$. Namely, 
the value of $\a_1$ is well determined from the slope.

In Fig.~\ref{Fig-screening}, 
the solid and dot-dash lines are lower than the threshold mass 
$2M_{j=1,n=0,\alpha=0}$ (dotted line). This guarantees 
that the $M_{\pi, {\rm scr}}$ and $M_{a_0, {\rm scr}}$ determined 
from the pole location in the complex-$\tilde{q}$ plane 
agree with those from the exponential decay 
of $\eta_{\xi\xi}(r)$ at large $r$.

\begin{figure}[t]
\begin{center}
   \includegraphics[width=0.4\textwidth]{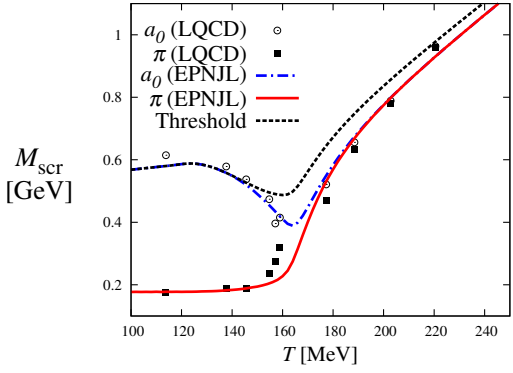}
\end{center}
\caption{$T$ dependence of pion and $a_0$-meson screening masses, 
$M_{\pi,{\rm scr}}$ and $M_{a_0,{\rm scr}}$. 
The solid (dot-dash) line denotes 
$M_{\pi,{\rm scr}}$ ($M_{a_0,{\rm scr}}$) calculated by 
the EPNJL model, whereas 
the dotted line corresponds to the threshold mass. 
LQCD data are taken from Ref.~\cite{Cheng:2010fe}; 
closed squares (open circles) correspond to 
the 2+1 flavor data 
for $M_{\pi,{\rm scr}}$ ($M_{a_0,{\rm scr}}$). 
In Ref. \cite{Cheng:2010fe}, 
$T_{c}$ was considered $196~{\rm MeV}$, 
but it was refined to $154\pm 9$ MeV 
\cite{Borsanyi:2010bp,Bazavov:2011nk}. 
The latest value is taken in this figure. 
}
\label{Fig-screening}
\end{figure}

\begin{figure}[t]
\begin{center}
 \includegraphics[width=0.4\textwidth]{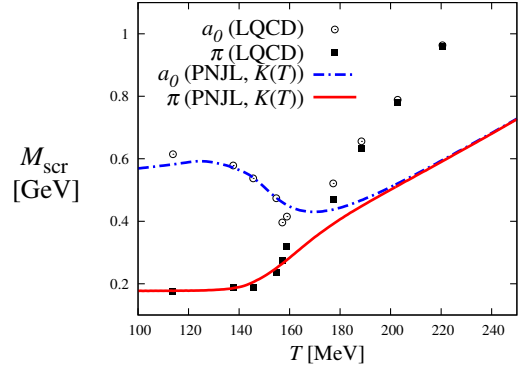}
\end{center}
\caption{Effects of $T$-dependent KMT interaction on 
pion and $a_0$-meson screening masses. 
The solid (dot-dash) line denotes 
$M_{\pi,{\rm scr}}$ ($M_{a_0,{\rm scr}}$) calculated by 
the PNJL model with $T$-dependent coupling $K(T)$. 
See Fig. \ref{Fig-screening} for LQCD data.  
}
\label{Fig-screening-K(T)-PNJL}
\end{figure}

In the EPNJL model with the present parameter, the chiral susceptibility 
$\chi_{ll}$ for light quarks has a peak at $T=163$ MeV, 
as shown later in Fig. \ref{Fig-condensate-chiral-limit}(a). 
This indicates $T_c=163$ MeV.  
The model result is consistent with LQCD data 
$T_{\rm c}= 154\pm 9$~MeV of Refs. \cite{Borsanyi:2010bp,Bazavov:2011nk}. 
For the deconfinement transition, meanwhile, 
the parameter $T_0$ is adjusted to reproduce LQCD data 
on $T_c^{\rm deconf}$, as already mentioned in Sec. \ref{Model setting}.  
In fact, the Polyakov-loop susceptibility $\bar{\chi}_{\Phi {\bar \Phi}}$ 
has a peak at $T=165$ MeV in the EPNJL model, as shown in Fig. 
\ref{Fig-condensate-chiral-limit}(b). 
The model result $T_c^{\rm deconf}=165$ MeV is consistent 
with LQCD data $T_c^{\rm deconf}=170 \pm 7$ MeV of Ref. \cite{Aoki:2009sc}.

\begin{figure}[t]
\begin{center}
   \includegraphics[width=0.4\textwidth]{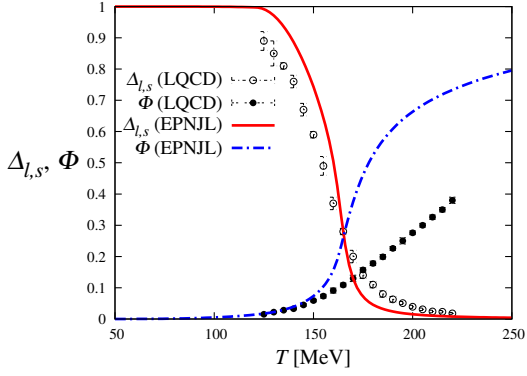}
\end{center}
\caption{$T$ dependence of $\Delta_{l,s}$ and $\Phi$. 
The solid (dot-dash) line corresponds to results of the EPNJL model 
for $\Delta_{l,s}$ ($\Phi$). 
LQCD data for 2+1 flavors are taken from 
Ref.~\cite{Borsanyi:2010bp}. 
}
\label{Fig-condensate}
\end{figure}

Figure~\ref{Fig-condensate} shows $T$ dependence of 
the renormalized chiral condensate $\Delta_{l,s}$ defined by
\beq
\Delta_{l,s}\equiv\frac{\sigma_l(T)-\frac{m_l}{m_s}\sigma_s(T)}
{\sigma_l(0)-\frac{m_l}{m_s}\sigma_s(0)},
\eeq
and the Polyakov loop $\Phi$. 
The present EPNJL model well reproduces LQCD data~\cite{Borsanyi:2010bp} 
for the magnitude of $\Delta_{l,s}$ in addition to the value of $T_c$. 
The present model overestimates LQCD data for the magnitude of 
$\Phi$, although it yields a result consistent with LQCD 
for $T_c^{\rm deconf}$. 
The overestimation in the magnitude of $\Phi$ is a famous problem 
in the PNJL model. 
Actually, many PNJL calculations have this overestimation. 
This is considered to come from the fact that 
the definition of the Polyakov loop is different between LQCD 
and the PNJL model~\cite{Braun:2007bx,Marhauser:2008fz}. 
In LQCD the definition is 
$\Phi_{\rm LQCD} 
=\langle {\rm tr}_{{\rm c}}~\mathcal{P}\exp[i\int_{0}^{1/T} d\tau A_4(\tau,{\bf x})] \rangle/3$, 
while in the PNJL model based on the Polyakov gauge and 
the mean field approximation the definition is $\Phi_{\rm PNJL}
= {\rm tr_c} \exp[i \langle A_4 \rangle/T] /3$, 
although both are order parameters of $Z_3$ 
symmetry~\cite{Braun:2007bx,Marhauser:2008fz}; see for example Ref.~\cite{Megias,Arriola:2014} as a trial to solve this
problem.

\begin{figure}[t]
\begin{center}
 \includegraphics[width=0.4\textwidth]{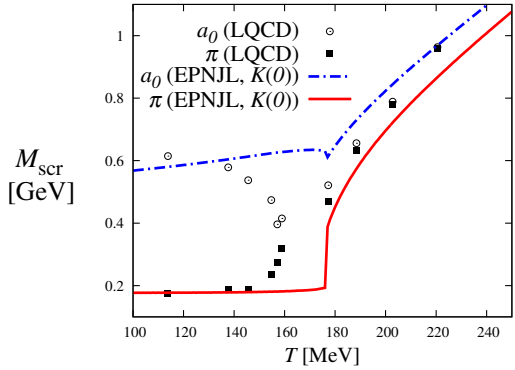}
\end{center}
\caption{Effects of $T$-dependent KMT interaction on 
pion and $a_0$-meson screening masses. 
The solid (dot-dash) line denotes 
$M_{\pi,{\rm scr}}$ ($M_{a_0,{\rm scr}}$) calculated by 
the EPNJL model with $K(T)=K(0)$. 
See Fig. \ref{Fig-screening} for LQCD data.  
}
\label{Fig-screening-K(0)}
\end{figure}

Now we investigate effects of $T$-dependent KMT interaction $K(T)$ on 
$M_{\pi,{\rm scr}}$ and $M_{a_0,{\rm scr}}$. 
In Fig. \ref{Fig-screening-K(0)}, $T$-dependence of $K(T)$ is switched off; 
namely, results of the EPNJL model with $K(T)=K(0)$ are shown. 
One can see that 
$T$-dependence of $K(T)$ reduces the mass difference 
between $M_{\pi,{\rm scr}}$ and $M_{a_0,{\rm scr}}$
significantly in a range $150 \la T \la 180$~MeV, 
comparing Fig. \ref{Fig-screening-K(0)} with Fig. \ref{Fig-screening}. 
At $T = 176$ MeV where 
first-order chiral and deconfinement transitions take place, 
$M_{\pi,{\rm scr}}$ has a jump while $M_{a_0,{\rm scr}}$ has a cusp. 
Meson screening mass is thus a good indicator 
for a first-order transition.

\begin{figure}[t]
\begin{center}
 \includegraphics[width=0.4\textwidth]{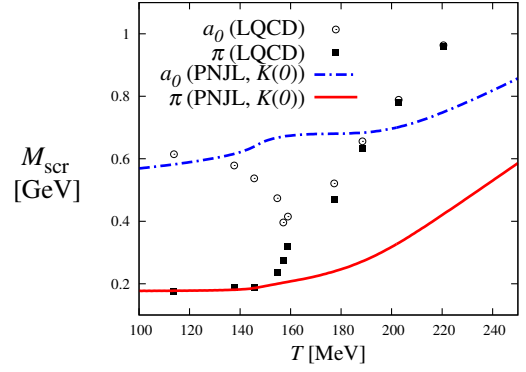}
\end{center}
\caption{Effects of the entanglement coupling 
$G_{\rm s}(\Phi)$ on pion and $a_0$-meson screening masses. 
The solid (dot-dash) line denotes 
$M_{\pi,{\rm scr}}$ ($M_{a_0,{\rm scr}}$) calculated by the standard PNJL model with constant $K$, i.e., the EPNJL model with $K(T)=K(0)$ and $\a_1=0$.
See Fig. \ref{Fig-screening} for LQCD data.  
}
\label{Fig-screening-K(0)-PNJL}
\end{figure}

In Fig. \ref{Fig-screening-K(0)-PNJL}, 
both the $T$ dependence of $K(T)$ and the entanglement of 
$G_{\rm s}(\Phi)$ are switched off. Namely, the results of 
the standard PNJL model with a constant $K$ are shown.  
The model cannot reproduce LQCD data, as expected.

\begin{figure}[t]
\begin{center}
   \includegraphics[width=0.4\textwidth]{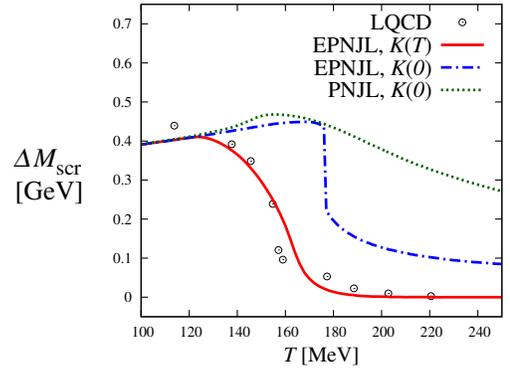}
\end{center}
\caption{Mass difference $\Delta M_{\rm scr}(T)$ 
between pion and $a_0$-meson screening masses. 
The solid, dot-dash and dotted lines denote results of 
the EPNJL model, the EPNJL model with $K(T)=K(0)$ and 
the standard PNJL model with $K(T)=K(0)$, respectively. 
See Fig. \ref{Fig-screening} for LQCD data.  
}
\label{Fig-screening-difference}
\end{figure}

Figure \ref{Fig-screening-difference} shows three types of 
EPNJL calculations for the mass difference $\Delta M_{\rm scr}(T)$. 
The mass difference plays a role of the order parameter of 
the effective restoration of $U(1)_{\rm A}$. 
The full-fledged EPNJL calculation (solid line) 
with both $T$-dependent $K$ and 
the entanglement coupling $G_{\rm s}(\Phi)$ well 
reproduces LQCD data, while the standard PNJL model 
(dotted line) with constant $K$ largely overestimates the data. 

The present model has $T$ dependence implicitly in $G_{\rm s}(\Phi)$   
through $\Phi$ and explicitly in $K(T)$. 
As a model opposite to the present one, one may consider 
the case that $K(T)=K(0)$ and $G_{\rm s}$ has $T$ dependence explicitly,
i. e., $G_{\rm s}=G_{\rm s}(T)$. We can determine $G_{\rm s}(T)$ so as to reproduce LQCD data for $\Delta_{l,s}$, however, this model overestimates LQCD data for $\Delta M_{\rm scr}$. 
Thus the present model is well designed.

\subsection{Meson susceptibilities}
\label{Meson susceptibilities}

The validity of $K(T)$ is investigated by comparing 
LQCD data with the  model results for 
meson susceptibilities $\chi_{\xi}^{\rm sus}$ ($\xi=\pi,a_0,\eta_{\rm ns}, 
\sigma_{\rm ns}$). 
LQCD data based on domain-wall fermions~\cite{Bhattacharya:2014ara}  
are available for two cases of pion mass $M_\pi(0) $ at vacuum 
being the physical value $135$ MeV and a slightly heavier value $200$ MeV. 
In order to reproduce these values with the EPNJL model, 
we take $m_l=5.68$ MeV for the first case and $12.8$ MeV for the second one.

\begin{figure}[t]
\begin{center}
   \includegraphics[width=0.4\textwidth]{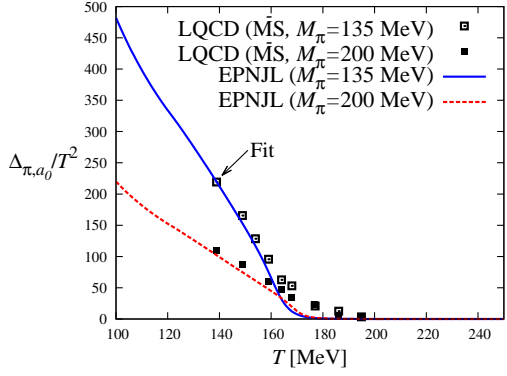}
\end{center}
\caption{$T$ dependence of the difference $\Delta_{\pi,a_0}$ 
between $\pi$ and $a_0$ meson susceptibilities 
for two cases of $M_\pi(0)=135$ and $200$ MeV.
}
\label{Fig-meson-susceptibility-U1A}
\end{figure}

We consider the difference   
$\Delta_{\pi,a_0}=\chi_{\pi}^{\rm sus}-\chi_{a_0}^{\rm sus}$ 
as an order parameter of the effective $U(1)_{\rm A}$-symmetry restoration.  
Figure \ref{Fig-meson-susceptibility-U1A} 
shows $T$ dependence of $\Delta_{\pi,a_0}/T^2$ for two cases of 
$M_\pi(0)=135$ and $200$ MeV. 
Since the $\chi_{\xi}^{\rm sus}$ have ultraviolet divergence, 
they are renormalized with the $\overline{\rm MS}$ scheme in LQCD. 
For this reason, one cannot compare the LQCD data with the results 
of the EPNJL model directly. We then multiply the model results by a constant 
so as to reproduce LQCD data at $T=139$ MeV 
for the case of $M_\pi(0)=135$ MeV. 
The model results thus renormalized well reproduce 
LQCD data for any $T$ in both cases of $M_\pi(0)=135$ and $200$ MeV. 

\begin{figure}[t]
\begin{center}
   \includegraphics[width=0.4\textwidth]{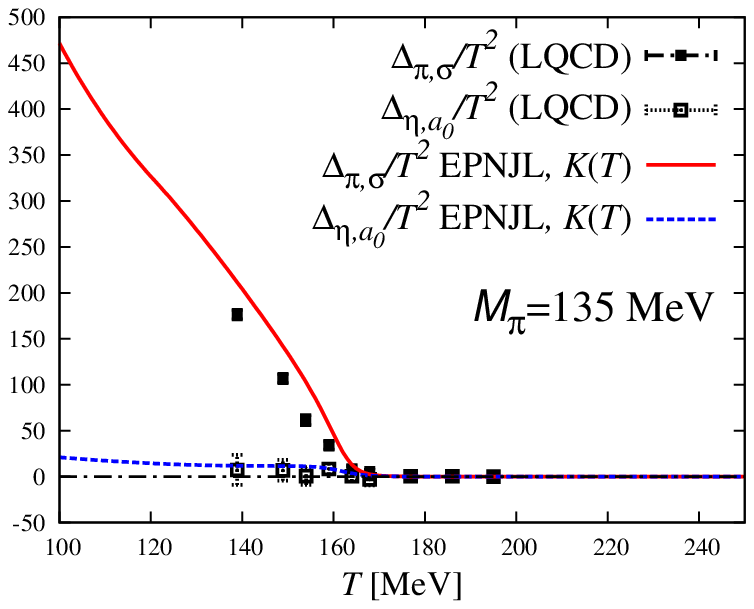}
\end{center}
\caption{$T$ dependence of $\Delta_{\pi,\sigma}$ and 
$\Delta_{\eta,a_0}$ for $M_\pi(0)=135$ MeV. 
}
\label{Fig-meson-susceptibility-chiral-1}
\end{figure}

\begin{figure}[t]
\begin{center}
   \includegraphics[width=0.4\textwidth]{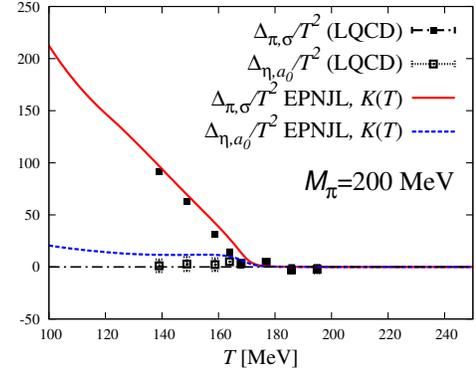}
\end{center}
\caption{$T$ dependence of $\Delta_{\pi,\sigma}$ and $\Delta_{\eta,a_0}$ for $M_\pi(0)=200$ MeV.
}
\label{Fig-meson-susceptibility-chiral-2}
\end{figure}

A similar analysis is made for $T$ dependence of 
$\Delta_{\pi,\sigma}=\chi_{\pi}^{\rm sus}-\chi_{\sigma_{\rm ns}}^{\rm sus}$ 
and 
$\Delta_{\eta,a_0}=\chi_{\eta_{\rm ns}}^{\rm sus}-\chi_{a_0}^{\rm sus}$ that 
are related to $SU_{\rm L}(2)\times SU_{\rm R}(2)$ symmetry. 
Figure \ref{Fig-meson-susceptibility-chiral-1} 
(\ref{Fig-meson-susceptibility-chiral-2}) shows $T$ dependence of
$\Delta_{\pi,\sigma}/T^2$ and $\Delta_{\eta,a_0}/T^2$ for 
$M_\pi(T)=135$ ($200$) MeV. In both the figures, the EPNJL model well 
reproduces $T$ dependence of LQCD results. The present model 
with the $K(T)$ of \eqref{T-dependent-K}  is thus reasonable.

\subsection{The order of chiral transition near the physical point}
\label{The order of chiral transition}

Finally we consider the order of chiral transition near the physical point 
{\bf
$(m_l^{\rm phys},m_s^{\rm phys})=(6.2{[\rm MeV]},175{[\rm MeV]})$} in the $m_l$--$m_s$ plane, First we vary $m_l$ from 9.9 to 0 MeV with $m_s$ fixed 
at 175~MeV. 

\begin{figure}[t]
\begin{center}
   \includegraphics[width=0.4\textwidth]{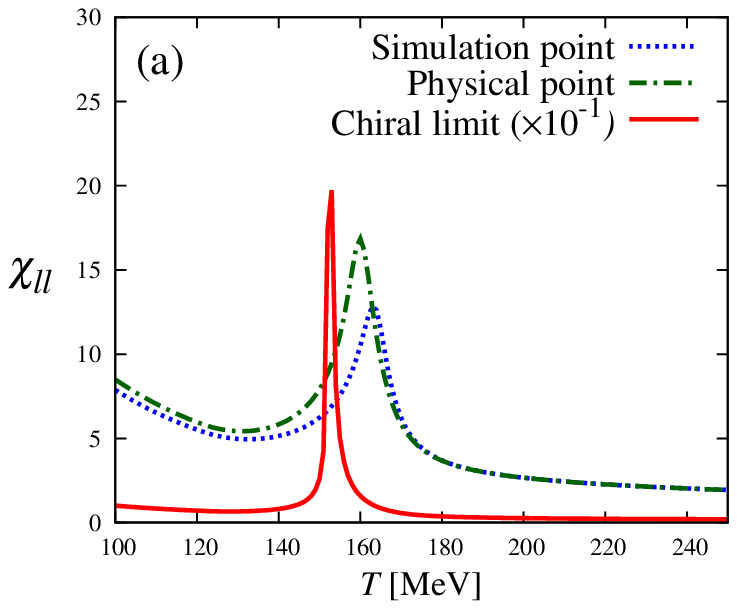}
   \includegraphics[width=0.4\textwidth]{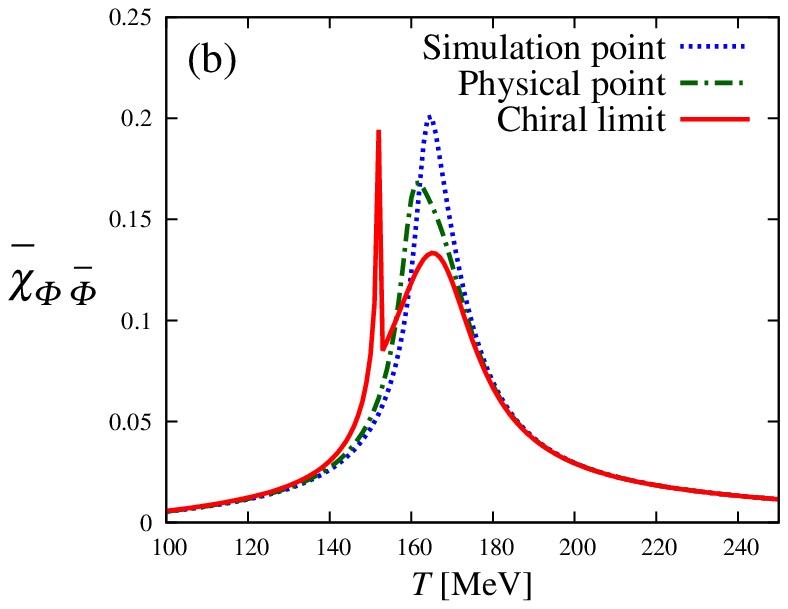}
\end{center}
\caption{
$T$ dependence of (a) chiral susceptibility $\chi_{ll}$ and 
(b) Polyakov-loop susceptibility $\bar{\chi}_{\Phi {\bar \Phi}}$ 
at S-point, P-point and C$_l$-point. 
Here $\chi_{ll}$ and $\bar{\chi}_{\Phi {\bar \Phi}}$ are dimensionless and 
their definition is the same as in the LQCD formulation. 
Calculations are done by 
the EPNJL model with the present parameter set. 
The dotted, dot-dash and solid lines stand for the results  
at S-point, P-point and C$_l$-point, respectively. 
At C$_l$-point, $\chi_{ll}$ is divided by 10 and diverges 
at $T=T_c=153$~MeV
}
\label{Fig-condensate-chiral-limit}
\end{figure}

Figure \ref{Fig-condensate-chiral-limit} 
presents $T$ dependence of the chiral susceptibility $\chi_{ll}$ 
for light quarks and the Polyakov-loop susceptibility 
$\bar{\chi}_{\Phi {\bar \Phi}}$ in three points, 
``simulation point (S-point)'' of $(m_l,m_s)=(9.9{[\rm MeV]},175{[\rm MeV]})$, 
``physical point (P-point)'' of $(m_l,m_s)=(6.2{[\rm MeV]},175{[\rm MeV]})$ 
and ``light-quark chiral-limit point (C$_l$ point)'' of 
$(m_l,m_s)=(0{[\rm MeV]},175{[\rm MeV]})$. 
In general, $T_c$ and $T_c^{\rm deconf}$ 
determined from peak positions of $\chi_{ll}$ and $\bar{\chi}_{\Phi {\bar \Phi}}$ 
depend on $m_l$ and $m_s$. However, as shown in panel (a), 
the $T_c$ thus determined is $163$~MeV at 
S-point and $160$~MeV at P-point, 
and hence the value little varies between the two points. 
At C$_l$ point, $\chi_{ll}$ diverges at $T=T_c=153$~MeV. 
The chiral transition is thus second order at C$_l$-point at least 
in the mean-field level. This result suggests that 
the effective $U(1)_{\rm A}$ restoration is not completed at $T=T_c$. 
This suggestion is supported by LQCD data at S-point in Fig. 
\ref{Fig-screening-difference} where 
$\Delta M_{\rm scr}(T_c)$ is about a half of $\Delta M_{\rm scr}(0)$.

As shown in panel (b), $m_l$ dependence of $T_c^{\rm deconf}$ is even smaller; 
namely, $T_c^{\rm deconf}=165$~MeV for S-point and C$_l$-point and 163~MeV 
for P-point. At C$_l$-point, $\bar{\chi}_{\Phi {\bar \Phi}}$ has a sharp peak 
at $T=153$~MeV. It is just a result of the propagation of divergence 
from $\chi_{ll}$ to $\bar{\chi}_{\Phi {\bar \Phi}}$ \cite{Kashiwa:2009zz}, 
and never means 
that a second-order deconfinement takes place there.

\begin{figure}[t]
\begin{center}
   \includegraphics[width=0.45\textwidth]{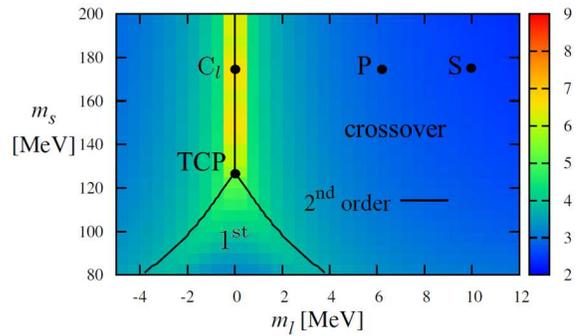}
\end{center}
\caption{
Order of chiral transition near physical point in the 
$m_l$--$m_s$ plane. 
The value of $\log[\chi_{ll}(T_c)]$ is shown by a change in hue. 
Simulation point, physical point, light-quark chiral-limit point and 
tricritical point are denoted by S, P, C$_l$ and TCP. 
The solid lines stand for second-order chiral transitions. 
}
\label{Fig-Columbia-plot}
\end{figure}

Next, both $m_l$ and $m_s$ are varied near P-point. 
Figure \ref{Fig-Columbia-plot} shows the value of 
$\log[\chi_{ll}(T_c)]$ near P-point in the $m_l$--$m_s$ plane. 
The value is denoted by a change in hue. 
Three second-order chiral transitions (solid lines) 
meet at $(m_l^{\rm tric},m_s^{\rm tric}) 
\approx (0,0.726m_s^{\rm phys}) = (0{[\rm MeV]},127{[\rm MeV]})$. 
This is a tricritical point (TCP) of chiral phase transition. 

\section{Summary}
\label{Summary}

In summary, 
we incorporated the effective restoration of $U(1)_{\rm A}$ symmetry 
in the 2+1 flavor EPNJL model by introducing a $T$-dependent coupling 
strength $K(T)$ to the KMT interaction. 
The $T$ dependence was well determined from 
state-of-the-art 2+1 flavor LQCD data 
on pion and $a_0$-meson screening masses. 
To derive the meson screening masses in the EPNJL model, 
we extended our previous prescription of Ref.~\cite{Ishii:2013kaa} 
for 2 flavors to 2+1 flavors. The strength $K(T)$ thus obtained is
suppressed in the vicinity of the pseudocritical temperature of chiral
transition.
As a future work, it is quite interesting to clarify 
how the present suppression is explained by instantons. 

 In order to check the validity of $K(T)$, we analyze 
$\pi,a_0, \eta_{\rm ns},\sigma_{\rm ns}$-meson susceptibilities 
obtained by state-of-the-art LQCD simulations with domain-wall 
fermions~\cite{Bhattacharya:2014ara}. The EPNJL model with 
the $K(T)$ of \eqref{T-dependent-K}
well reproduces $T$ dependence of LQCD data. 
The present model building is thus reasonable.

Using the EPNJL model with the present parameter set, we showed that, 
at least in the mean field level, 
the order of chiral transition is second order 
at the light-quark chiral-limit point 
of $m_l=0$ and $m_s=175$ MeV (the physical value). 
This result indicates that there exists a tricritical point 
near the light-quark chiral-limit point in the $m_l$--$m_s$ plane. 
We then estimated the location of the tricritical point as 
$(m_l,m_s) \approx (0{[\rm MeV]},127{[\rm MeV]})$.

In conclusion, we present a simple method for calculating 
meson screening masses in PNJL-like models. This allows us to compare model 
results with LQCD data on meson screening masses. 
Meson screening masses are quite useful to determine model parameters. 
In particular, the mass difference between 
pion and $a_0$-meson is effective to determine $T$ dependence of the KMT 
interaction. 
The EPNJL model with the present parameter set is useful 
for estimating the order of chiral transition at 
the light-quark chiral-limit point and the location of the tricritical 
point, since it is hard to reach the chiral regime directly with LQCD.

The present model consists of $\Phi$-dependent four-quark interactions 
and $T$-dependent six-quark interactions. Meanwhile, the importance of 
eight-quark interactions was pointed out in 
Ref.~\cite{Osipov:2006}, since it makes the thermodynamic potential bounded from 
below. Furthermore, it is reported in Ref.~\cite{Osipov:2013} that current-quark-mass 
dependence of quark-quark interactions 
is effective to reproduce meson pole masses with 
good accuracy. Therefore, further
inclusion of these interactions 
is interesting as a future work. 

\noindent
\begin{acknowledgments}
M. I, J. T., H. K., and M. Y. are supported by Grant-in-Aid for Scientific Research (
No. 27-3944, No. 25-3944, No. 26400279 and No. 26400278) from the Japan Society for the Promotion of Science (JSPS). 
\end{acknowledgments}


\end{document}